\documentclass{aa}
\usepackage{epsfig}

\begin{document}

\thesaurus{10.08.1, 10.11.1, 10.19.2, 12.04.1, 12.07.1}

\title{Microlensing towards the Small Magellanic Cloud\\
EROS 2 first year survey
\thanks{Based on observations made at the European Southern Observatory,
La Silla, Chile.}}
\author{
N.~Palanque-Delabrouille\inst{1,9}, 
C.~Afonso\inst{1}, 
J.N.~Albert\inst{2},
J.~Andersen\inst{6},
R.~Ansari\inst{2}, 
\'E.~Aubourg\inst{1}, 
P.~Bareyre\inst{1,4}, 
F.~Bauer\inst{1},
J.P.~Beaulieu\inst{5},
A.~Bouquet\inst{4},
S.~Char\inst{7},
X.~Charlot\inst{1},
F.~Couchot\inst{2}, 
C.~Coutures\inst{1}, 
F.~Derue\inst{2}, 
R.~Ferlet\inst{5},
J.F.~Glicenstein\inst{1},
B.~Goldman\inst{1,10,11},
A.~Gould\inst{1,8},
D.~Graff\inst{1},
M.~Gros\inst{1}, 
J.~Haissinski\inst{2}, 
J.C.~Hamilton\inst{4},
D.~Hardin\inst{1},
J.~de Kat\inst{1}, 
\'E.~Lesquoy\inst{1},
C.~Loup\inst{5},
C.~Magneville \inst{1}, 
B.~Mansoux\inst{2}, 
J.B.~Marquette\inst{5},
\'E.~Maurice\inst{3}, 
A.~Milsztajn \inst{1},  
M.~Moniez\inst{2},
O.~Perdereau\inst{2},
L.~Pr\'evot\inst{3}, 
C.~Renault\inst{1},  
J.~Rich\inst{1}, 
M.~Spiro\inst{1},
A.~Vidal-Madjar\inst{5},
L.~Vigroux\inst{1},
S.~Zylberajch\inst{1}
\\   \indent   \indent
The EROS collaboration
}
%% 1 Saclay, 2 LAL, 3 Marseille, 4 IAP, 5 Copenhague, 6 La Serena 7 Gould
\institute{
CEA, DSM, DAPNIA,
Centre d'\'Etudes de Saclay, F-91191 Gif-sur-Yvette Cedex, France
\and
Laboratoire de l'Acc\'{e}l\'{e}rateur Lin\'{e}aire,
IN2P3 CNRS, Universit\'e Paris-Sud, F-91405 Orsay Cedex, France
\and
Observatoire de Marseille,
2 place Le Verrier, F-13248 Marseille Cedex 04
\and
Coll\`ege de France, Laboratoire de Physique Corpusculaire, IN2P3 CNRS, 
11 place Marcellin Berthelot, F-75231 Paris Cedex, France
\and
Institut d'Astrophysique de Paris, INSU CNRS,
98~bis Boulevard Arago, F-75014 Paris, France
\and
Astronomical Observatory, Copenhagen University, Juliane Maries Vej 30, 
DK-2100 Copenhagen, Denmark
\and
Universidad de la Serena, Facultad de Ciencias, Departamento de Fisica,
Casilla 554, La Serena, Chile
\and
Department of Astronomy, Ohio State University, Columbus, OH 43210, U.S.A.
\and
Enrico Fermi Institute, University of Chicago, 5460 South Ellis Avenue, 
Chicago, IL 60637, U.S.A.
\and
Dept. Astronom\'ia, Universidad de Chile, Casilla 36-D, Santiago,
Chile
\and
European Southern Observatory, Casilla 19001, Santiago 19, Chile
}
\offprints{Nathalie.Delabrouille@cea.fr}

\date{Received;accepted}

\authorrunning{N. Palanque-Delabrouille et al.}
\titlerunning{Microlensing towards the SMC --- EROS 2 first year survey}

\maketitle
\begin{abstract}
We present here an analysis of the light curves of 5.3 million stars in the
Small Magellanic Cloud observed by EROS (Exp\'erience de Recherche d'Objets
Sombres). One star exhibits a variation that is best interpreted as due to
gravitational microlensing by an unseen object. This candidate was also
reported by the MACHO collaboration. Once corrected for blending, the Einstein
radius crossing time is 123 days, corresponding to lensing by a Halo object of
$2.6^{+8.2}_{-2.3} {\rm \;M_{\odot}}$. The maximum magnification is a factor of
2.6. The light curve also displays a periodic modulation with a 2.5\% amplitude
and a period of 5.1~days.  Parallax analysis of the candidate indicates that a
Halo lens would need to have a mass of at least $0.3 {\rm \; M_{\odot}}$,
although a lens in the SMC could have a mass as low as $0.07 {\rm\;
M_{\odot}}$. We estimate the optical depth for microlensing towards the SMC due
to this event to be $\sim 3.3 \times 10^{-7}$, with an uncertainty dominated by
Poisson statistics. We show that this optical depth corresponds to about half
that expected for a spherical isothermal Galactic Halo comprised solely of such
objects, and that it is consistent with SMC self-lensing if the SMC is
elongated along the line-of-sight by at least 5 kpc.

\end{abstract}

\section{Introduction}

Ten years after Paczy\'{n}ski's proposal (\cite{Paczynski}) to use
gravitational microlensing as a tool for discovering dark stars, and four years
after the identification of the first candidate events in the direction of the
Large Magellanic Cloud (LMC) (Alcock et al. 1993, Aubourg et al. 1993) and
Galactic Bulge (\cite{Ogle}), searches for microlensing events have started to
yield quantitative information that contributes to a better understanding of
Galactic structure (Stanek et al. 1996).  Probably the most intriguing result
is that the measured optical depth for microlensing towards the LMC implies a
total Galactic Halo mass in compact objects that is within a factor of two of
that required to explain the rotation curves of spiral galaxies
(\cite{Macho2yr}, see also Ansari et al. 1996a). The time scales associated
with these events indicate surprisingly high mass lenses and the difficulty of
accounting for the events with known stellar populations has stimulated
interest in star formation and evolution processes. Strong limits have been set
on the maximum contribution of low mass objects to the Halo of the Milky Way
(Renault et al. 1997, see also Alcock et al. 1996).

Given the importance of these results, it is imperative to verify them by using
other lines of sight, the most promising ones being the Small Magellanic Cloud
(SMC) and M31. Here, we present a first analysis of microlensing data in the
direction of the SMC by using 5.3 million light curves collected by EROS2
during the first year of the survey. More details can be found in
(Palanque-Delabrouille 1997).

\section{Experimental setup}

Our results have been obtained with a completely redesigned setup. The EROS
program now uses exclusively the dedicated 1 meter MARLY telescope, specially
refurbished and fully automated for the EROS2 survey (Bauer et al. 1997), now
in operation at the European Southern Observatory at La Silla, Chile. The
telescope optics allows simultaneous imaging in ``blue'' ($\lambda {\rm\;in\;}
420-720$~nm, peak at $\lambda \simeq 560$~nm) and ``red'' ($\lambda {\rm\;in\;}
620-920$~nm, peak at $\lambda \simeq 760$~nm) wide pass-bands of a
one-square-degree field. This is achieved by a beam-splitting dichroic cube
with a CCD camera mounted behind each channel. Each camera contains a mosaic of
8 Loral 2048 x 2048 thick CCD's. The total field is 0.7~deg (right ascension) x
1.4~deg (declination). The pixel size is 0.6 arcsec, and typical global image
quality (atmospheric seeing + instrument) is  2 arcsec FWHM.

The read-out of the entire mosaic is done in parallel, controlled by Digital
Signal Processors, and takes 50 seconds. The data are first transferred to two
VME crates (one per color), which manage the real-time part of the acquisition
system, and then to two Alpha workstations where a quality assessment is run
(monitoring CCD defects, sky background, seeing, number of stars\ldots) and
flat-field reduction is done. The raw and reduced data are finally saved on DLT
tapes.

\begin{figure} [h] \begin{center} 
  \epsfig{file=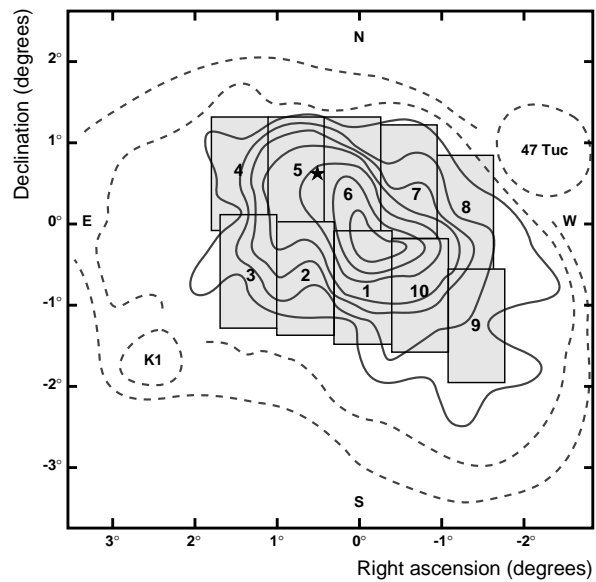,width=7.8cm} \vspace{-.1cm} \caption{Position of
  the SMC fields (sm001 to sm010) on the de Vaucouleurs' isophote map (de
  Vaucouleurs 1957). The star in field 5 indicates the position of the
  microlensing candidate. The isophote levels (inner to outer ones) are 21.4,
  21.7, 22.0, 22.5, 22.8, 23.2, 23.9 ${\rm mag/arcsec^2}$ in red light
  ($\lambda = 625$nm).}  \label{carteSMC}\end{center}\vspace{-0.5cm}
\end{figure}

Data taking with the new apparatus began in July, 1996.  Microlensing targets
include fields near the galactic center, in the disk of the Galaxy, and in the
LMC and SMC. The data discussed here concern 10 fields covering the densest
$10\;{\rm deg}^{2}$ of the SMC, as illustrated in figure \ref{carteSMC}.
The fields were observed from July, 1996 to February, 1997 and
again starting in July, 1997. During the 1996-97 season, from 60 to 120 usable
images were taken of each field, giving a sampling time of one point every 2-4
days on average. Exposure times varied from 5~min in the central fields to
15~min in the outermost fields.

The DLT tapes produced in Chile are shipped to the CCPN (IN2P3 computing
center, CNRS) in Lyons, France, where data processing occurs.  For each of the
fields, a template image is first constructed by adding together 10 exposures
of good quality, each re-sampled by a factor of 0.7. A reference star catalog
is then built using the {\sc corrfind} star finding algorithm (the stellar
detection is done on a correlation image, obtained as described in section 3,
page \pageref{corr}).  For each subsequent image, after geometrical alignment
to the template catalog, each star identified on the reference catalog is
fitted together with neighboring stars, using a PSF determined on bright
isolated stars and imposing the position from the reference catalog.  A
relative photometric alignment is then performed, assuming most stars do not
vary.  Photometric errors are computed for each measurement, assuming again
that most stars are stable, and parameterized as a function of star brightness
and image sequence number.  Photometric accuracy is in the $8-20\%$ range at
magnitude $V\sim 20$ (depending on image quality), and of the order of $2\%$
for bright stars ($V\sim 17$).\footnote{This analysis does not require absolute
calibration and the magnitudes mentioned here are only approximate. Absolute
calibration is in progress.}  The number of reconstructed stars varies from
$8\times 10^5 {\rm \: deg}^{-2}$ in the densest region (where errors are
dominated by crowding) to $4\times 10^5 {\rm \: deg}^{-2}$ in the outer regions
(where errors are dominated by signal-to-noise). The photometry is described in
more details in (\cite{Peida}).

\section{Data analysis}

The 5.3 million light curves are subjected to a series of selection criteria
and rejection cuts (globally called ``cuts'') to isolate microlensing
candidates (Palanque-Delabrouille 1997). The first three (1--3) make use of the
expected general characteristics of microlensing candidates: single variations
on otherwise constant light curves, which coincide in time for data taken in
both colors. The next two cuts (4 and 5) are designed to reject a known
background of variable stars, while the last two (6 and 7) improve the
signal-to-noise of the set of selected candidates. The criteria were
sufficiently loose not to reject events affected by blending or by the finite
size of the source, or events involving multiple lenses or sources. We define a
positive (negative) fluctuation as a series of data points that (i) starts by
one point deviating by at least $1\sigma$ from the base flux, (ii) stops with
at least three consecutive points below ${\rm base\:flux} + 1\sigma$ (above
${\rm base\:flux} - 1\sigma$) and (iii) contains at least 4 points above ${\rm
base\:flux} + 1\sigma$ (below ${\rm base\:flux} - 1\sigma$). The significance
$LP$ of a given variation is defined as the negative of the logarithm of the
product, over the data points it contains, of the probability that each point
deviates from the base flux by more than the observed fluctuation ($x_i$ is the
deviation of the point taken at time $t_i$, in $\sigma$'s, $N$ is the number of
points within the fluctuation):
\begin{equation}
 LP = - \sum_{i=1}^{i=N}
   \log\left(\frac{1}{2}\:{\rm Erfc}\left(\frac{x_{i}}{\sqrt{2}}\right)\right) 
\end{equation}
We order the fluctuations along a light curve by decreasing significance. The
cuts of the analysis are described hereafter:
\begin{itemize}
\item 1: The main fluctuation detected in the red and blue light curves should
be both positive and occur simultaneously: if I is the time interval during
which the data are more than $1\sigma$ away from the base flux, we require
$({\rm I_{red}} \cap {\rm I_{blue}})/({\rm I_{red}} \cup {\rm I_{blue}})>20\%$.
\item 2: To reject flat light curves with only statistical fluctuations, we
require that on a given light curve \\ 
$LP({\rm 2^{nd}\:most\:significant\:fluct.})\:/\:LP({\rm main\:fluct.})<0.35$\\
in both colors.\footnote{This does not reject multiple lenses with caustic
crossings since all the points magnified would be contained in the same main
fluctuation.}
\item 3: We require that $LP(\rm main\;fluct.)>30$ in both colors.
\item 4: To exclude short period variable stars which exhibit scattered light
curves, we require that the RMS of the distribution of the deviation, in
$\sigma$'s, of each flux measurement from the linear interpolation between its
two neighboring data points be smaller than 2.5.
\item 5: We remove two under-populated regions of the color-magnitude diagram
(see figure \ref{cut_hr}) that contain a large fraction of variable stars
($\beta$ Cephei, RV Tauri variables, semi-regular giant variables and Mira Ceti
stars), defined by (flux $F$ given for the EROS2 filters --- $R$ for red and
$B$ for blue --- normalized to an exposure time of 300 s): \\
$\log(F_R/F_B)<-0.20$ and $\log(F_R)>4.5$ \\ $\log(F_R/F_B)>+0.07$ and
$\log(F_R)>2.7$
\begin{figure} [h] \vspace{-.2cm} \begin{center} \vspace{8cm}
  \caption{Cut on
  color-magnitude diagram, here shown for field sm001. The dotted lines delimit
  the rejected areas. The dots correspond to all the light curves in the field,
  the star markers are the remaining objects for this field, after cuts 1
  through 4.}  \label{cut_hr}\end{center}\vspace{-0.5cm}
\end{figure}
\item 6: We remove events with low signal-to-noise by requiring a significant
improvement of a microlensing fit (ml) over a constant flux fit (cst),
i.e. that \\
$[\chi^2({\rm cst}) -\chi^2({\rm ml})]/[\chi^2({\rm ml})/{\rm d.o.f.}]>150$ \\
where d.o.f. is the number of degrees of freedom.
\item 7: We require that the maximum magnification in the microlensing fit be
greater than 1.40.
\end{itemize}

The tuning of each cut and the estimate of the efficiency of the analysis is
done with Monte Carlo simulated light curves. To ensure similar photometric
dispersion on simulated events and on the data, the events are added to {\em
real\,} light curves. The microlensing parameters are drawn uniformly in the
following intervals: time of maximum magnification $t_0 \in [t_{\rm first}-150,
t_{\rm last}+150]$ days, impact parameter normalized to the Einstein radius
$u_0 \in [0,2]$ and time-scale (Einstein radius crossing time) $\Delta t \in
[0,150]$ days. We correct for blending statistically, using a study of the
typical flux distribution of the source stars which contribute to the flux of a
reconstructed star, depending on its position in the color-magnitude diagram.

\begin{table}[ht]
\begin{center} \vspace{-0.1cm} \begin{tabular}{|l||c|c|c|} 
\hline
 &{\em Number of}& \multicolumn{2}{c|}{\em Fraction of remaining} \\
{\em Cut description}&{\em stars}&\multicolumn{2}{c|}{\em stars removed by cut}
\\ \cline{3-4}
& {\em remaining} & {\em \ \ \ Data\ \ \ } & {\em Simulation}\\
\hline\hline 
Stars analyzed & 5,277,858 & - & - \\
1: Simultaneity& 125,071 & 98\% & 80\%\\
2: Uniqueness & 36,032 & 71\% & 13\%  \\
3: Significance& 4,022  & 89\% & 13\% \\
4: Stability& 1,214 & 70\% & 16\% \\
5: HR diagram& 463 & 62\% & 6\% \\
6: Microlensing fit& 48 & 89\% & 20\%  \\
7: Magnification& 10 & 76\% & 16\%  \\
\hline
\end{tabular}
\caption{Impact of each cut on data and simulated events. Each fraction 
for cut $n$ refers to the stars remaining after cut $(n-1)$.}\label{cuts}
\vspace{-0.5cm} \end{center}
\end{table}
Table \ref{cuts} summarizes the impact of the cuts. The first requirement
removes 98\% of the data light curves which are just flat light curves. It also
removes a large fraction of the simulated events of too low amplitude (large
impact parameter) or short-duration events peaking well outside the
observational period $[t_{\rm first},t_{\rm last}]$.  The other cuts remove a
large fraction of remaining data light curves (background) while leaving, in
general, at least 75\% of the simulated light curves.

The efficiency of the analysis (cuts 1 through 7) for detecting real
microlensing events is determined from the set of simulated microlensing
events, taking into account the effect of blending. The efficiencies (in \%)
normalized to an impact parameter $u_0<1$ and an observing period $T_{\rm obs}$
of one year are summarized in table \ref{eff} for various Einstein radius
crossing times $\Delta t$ (in days).
\begin{table}[ht]
\vspace{-0.1cm} \begin{tabular}{|c||c|c|c|c|c|c|c|c|c|} \hline
$\Delta t$ & 7 & 22&  37 & 52&  67 & 82 & 97 & 112& 127\\ \hline
$\epsilon(\Delta t)$ &8& 16& 20& 22 & 24& 27 & 28& 29& 30 \\ \hline
\end{tabular} \vskip .1cm
\begin{tabular}{|c||c|c|c|c|c|c|}  \hline
$\Delta t$ & 150 & 300 &  500&  1000 & 1500 & 2000\\ \hline
$\epsilon(\Delta t)$ & 29 & 28 & 27 & 24 & 19& 17\\
\hline
\end{tabular} \caption{Efficiency (in \%) of the analysis (cuts 1--7) for 
various time-scales $\Delta t$ (in days), normalized to $u_0<1$ and $T_{\rm
obs}=1{\rm \: yr}$. We monitor $N_{\rm obs}=5.3\times 10^6$ stars. The
efficiency values for $\Delta t>150$~days are obtained by simulating events of
duration $\Delta t$ over a period $[t_{\rm first}-\Delta t, t_{\rm last}+\Delta
t]$~days.} \label{eff} \vspace{-0.5cm}
\end{table}

\section{Study of the candidates}

Of the 5.3 million light curves, ten events passed all cuts and were inspected
individually. Scanning of Monte Carlo events indicates that a negligible number
of remaining candidates would be rejected by visual inspection. Three of the
ten candidates exhibit new variations on their light curve when adding the
first data from the second year survey, and are probably recurrent variable
stars. Another event has its light curve affected by the appearance of a
neighboring object which is below our detection threshold in the template
image, but becomes very bright for a period of about 60 days. The object could
be a nova or even a microlensing on an unresolved star. This analysis, however,
uses only stars identified on the template image, so the light curve must be
rejected. Four more events have light curves incompatible with microlensing and
are probably due to other physical processes (one of them is a nova in the SMC
(\cite{nova})). Another event exhibits a very chromatic variation ($A_R=1.7$
and $A_B=1.2$). If this were due to blending (i.e. when both the magnified star
and an undetected star contribute to the total reconstructed flux), the star
undergoing the magnification would be of similar brightness as clump giant
stars but redder than clump giants by about 1.2 magnitude. It could not belong
to the SMC and a microlensing interpretation of this event is thus
unrealistic. Thus, 9 out of the 10 candidates are rejected by this
inspection. \\

\begin{figure} [h] \begin{center} 
  \epsfig{file=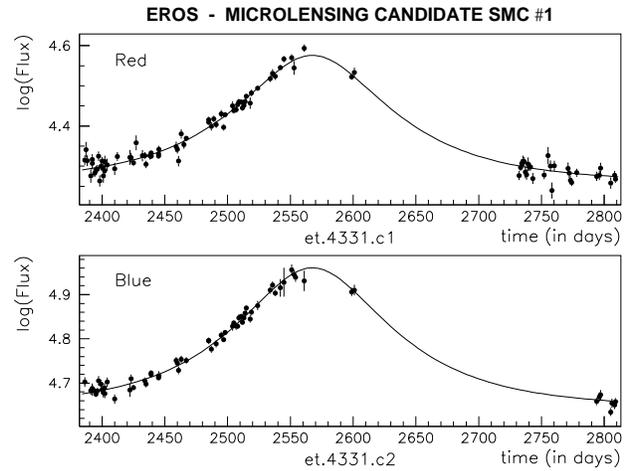,width=9cm} \caption{Light curve of microlensing
  candidate SMC \#1, with a standard microlensing fit (combined for the red and
  blue light curves) superimposed, with no blending assumed. Time is in days
  since Jan. 0, 1990 (Julian date 2,447,891.5).}
  \label{cdl_cand1}\end{center}\vspace{-0.5cm}
\end{figure}

The remaining light curve is shown in figure \ref{cdl_cand1}. It fits well the
standard microlensing hypothesis with an Einstein radius crossing time of 104
days, a maximum magnification of 2.1 (impact parameter $u_0=0.53$) occurring
on January 11, 1997 and $\chi^2/{\rm d.o.f.} = 268/161 = 1.7$.

The source star is located at $\alpha = 01^{\rm h}00'5.64''$ and $\delta =
-72^\circ 15'2.41''$ (J2000), and is labeled U0150\_00676152 in the USNO star
catalog. The magnitude of the source star given in the catalog is $R\simeq
18$. This microlensing candidate was also reported by the MACHO collaboration
and exhibited no variation during 3 years preceding the upward excursion
detected here (\cite{machosmc}).

Because of the high stellar density of the fields monitored in microlensing
surveys, the flux of each reconstructed star generally results from the
superposition of the fluxes of many source stars. We thus introduce two
additional parameters in the fit : the contribution $c_{\rm \,bl}$ in both
colors of the base flux of the magnified star ($f_{\rm star}$) to the total
base flux recovered ($f_{\rm star}+f_{\rm blend}$):
\begin{equation}
c_{\rm \,bl} \equiv \frac{f_{\rm star}}{f_{\rm star}+f_{\rm blend}}
             = \frac{A_{\rm reconstructed}-1}{A_{\rm real} - 1}
\end{equation}
The blending coefficient $c_{\rm bl}$ is unity when there is no blending and
$c_{\rm bl}\rightarrow 0$ in the limit where the magnified star does not
contribute at all to the total recovered baseline flux.  Allowing for blending
does not significantly improve the fit, but changes the best estimate values of
the fit parameters, as shown in table \ref{cand1}. The errors on the blending
coefficients are of 20\%.
\begin{table}[ht]
\begin{center} \vspace{-0.1cm} \begin{tabular}{|l||c|c|c|c|c|c|}
\hline
{\em Fit}&$u_0$&$t_0$&$\Delta t$&$c_{\rm \,bl}\;R$&$c_{\rm \,bl}\;B$&$\chi^2/{\rm d.o.f.}$ \\
\hline
{\em Red }& $0.53$ & $2567$ & $100$& - & - &164/94\\
{\em Blue }& $0.54$ & $2567$ & $103$& - & - &99/64\\
{\em Combined}& $0.53$&$2567$&$104$& - & - &268/161\\
{\em Blended}& $0.41$&$2567$&$123$&$0.71$&$0.71$&266/159\\
\hline
\end{tabular}
\caption{Results of microlensing fits to the SMC candidate. $t_0$ is the time
of maximum magnification and $\Delta t$ the Einstein radius crossing time, both
given in days.}\label{cand1}
\end{center} \vspace{-0.5cm}
\end{table}
The magnified source star would then have $\log(f_R) = 3.90$ and $\log(f_R/f_B)
= -0.38$ while the blend companion would have $\log(f_R) = 3.53$ and
$\log(f_R/f_B) = -0.39$. This amount of blending is in agreement with the
estimate given by the MACHO collaboration.  

The blending hypothesis is strengthened by the observation of a slight shift in
the position of the centroid of the reconstructed star as the magnification
occurs: $(\Delta \delta \simeq 0.072{\rm \;arcsec},\: \Delta \alpha \simeq
-0.090{\rm \;arcsec})$. This shift can be caused by the displacement of the
barycenter of the positions (weighted by the flux) of the two components of the
blend. The direction of this displacement (slope $|\Delta \alpha/\Delta
\delta|$ of $1.25$) is compatible with the direction of the apparent
elongation of the source star ($\Delta \delta \simeq -0.95{\rm \; arcsec},\:
\Delta \alpha \simeq 1.26 {\rm \; arcsec}$ i.e. a slope of 1.3) in the image of
the correlation coefficients between a Gaussian Point Spread Function (PSF) and
the template image (the correlation image is shown in figure \ref{candfits},
right):
\begin{equation}
{\rm coeff.} = \frac{{\rm covariance\,}(PSF, Image)}{\sqrt{{\rm variance\,}(P
SF)}\; \sqrt{{\rm variance\,}(Image)}} \label{corr}
\end{equation}
The correlation is calculated over a radius of $1.6\sigma$, where the $\sigma$
of the PSF is related to the seeing by ${\rm seeing} = 2\sqrt{2\ln 2}\,\sigma$.
The reconstructed source star might therefore consist of two components,
located approximately 1.5~arcsec apart. Both the template image and the
correlation image of an $8{\rm \;arcsec \times 8 \;arcsec}$ area around the
candidate are shown in figure \ref{candfits}. The pixel size on the template
image is 0.42~arcsec, that on the correlation image is 0.21~arcsec.
\begin{figure} [h] \begin{center} \vspace{6cm}
  \caption{Template image and
  correlation image around the candidate. North is to the left, East is up.}
  \label{candfits}\end{center}\vspace{-0.2cm}
\end{figure}
Note the clear improvement in stellar separation (and in stellar detection, as
evidenced with the leftmost star) in the correlation image as compared to the
template image, which allows us to infer the existence of a blend companion and
estimate its position. Requiring on the template image (figure \ref{candfits},
left) the existence of two source stars located along the observed position
angle (instead of a single star recovered as with the standard star finding
algorithm), we can estimate the flux ratio of the two stars. In the best fit,
the light is split with the ratio 70\% to 30\% between the two blended
components. This independent method thus gives a result consistent with that of
the microlensing fit.

\begin{figure} [h] \begin{center} \vspace{-.2cm}
  \epsfig{file=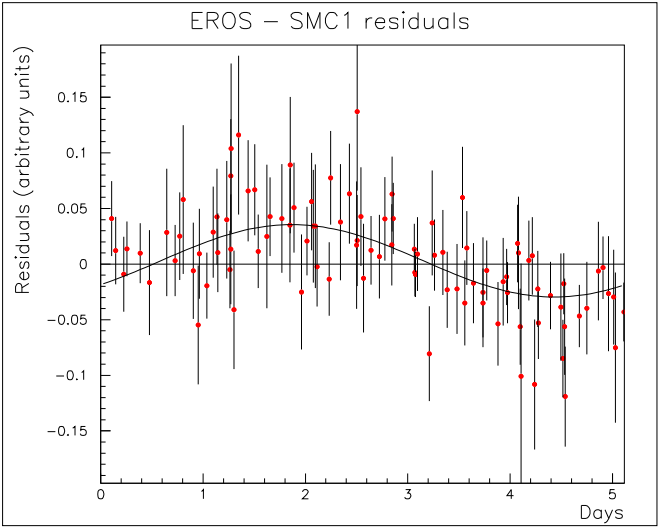,width=8.7cm} \caption{Light curve of the residuals
  in red, folded to $P = 5.124$~days.}
  \label{residus}\end{center}\vspace{-0.5cm}
\end{figure}

As can be seen in figure \ref{cdl_cand1}, the measured luminosities exhibit an
abnormally high scatter in the two colors.  Correspondingly, the $\chi^2 / {\rm
d.o.f.}$ (266/159 in Table 2) has a low probability, of order $10^{-5}$.  As a
significant correlation is observed between the residuals of the fit in both
colors (99.5\% CL), a search for periodicity was performed on these residuals.
The most likely period was found at $P = 5.124$~days, with a false detection
probability of $2.\:10^{-6}$ (other periods, aliases of $P$, are less probable,
though not excluded).  Figure \ref{residus} shows the residuals light curve,
folded to $P = 5.124$~days. About 1.5\% of the main sequence stars of similar
brightness have a light curve which exhibits a periodic modulation of at least
a $\sim$2\% amplitude.

We then repeated the microlensing fits of table 2, including a sinusoidal
modulation with three additional free parameters: period, phase and amplitude
of the modulation (identical for both colors).  Results of the fit are given in
table \ref{tabresidus}. The fits give almost identical results for $u_0, t_0,
\Delta t$ and blending factors as those in table 2 (respectively 0.42, 2568.,
123. and 0.74).  Of course, the modulation can affect either the amplified
component, or the blended companion.  The results of the fits favor the first
possibility, though only at the $2.5\sigma$ level.  We expect that the data
reported by the MACHO collaboration (Alcock et al., 1997b) have enough
statistical power to discriminate between these two possibilities.  We remark
that the $\chi^2$ of the fits including a modulation term are satisfactory,
indicating accurate modeling of errors.
\begin{table}[ht]
\begin{center} \vspace{-0.1cm} \begin{tabular}{|l||c|c|c|c|}
\hline
{\em Star modulated} & \begin{tabular}{cc} $A_{\rm mod.}$\\(in \%) 
   \end{tabular}& \begin{tabular}{cc} $P_{\rm mod.}$\\ (in days) \end{tabular} 
   & $\chi^2/{\rm d.o.f.}$ \\
\hline
{\em magnified star} & $2.9 \pm 0.5$ & $5.128 \pm 0.004$ & 157/156\\ 
{\em blend companion} & $11. \pm 7.$ & $5.128 \pm 0.004$ & 163/156 \\
\hline
\end{tabular}
\caption{Result of microlensing fit + sinusoidal modulation on either component
of the blend. $A_{\rm mod.}$ is the amplitude of the modulation normalized to
the unamplified stellar flux, and $P_{\rm mod.}$ the period.}
\label{tabresidus} \end{center} \vspace{-0.5cm}
\end{table}

\begin{figure} [h] \begin{center} \vspace{-.5cm}
  \epsfig{file=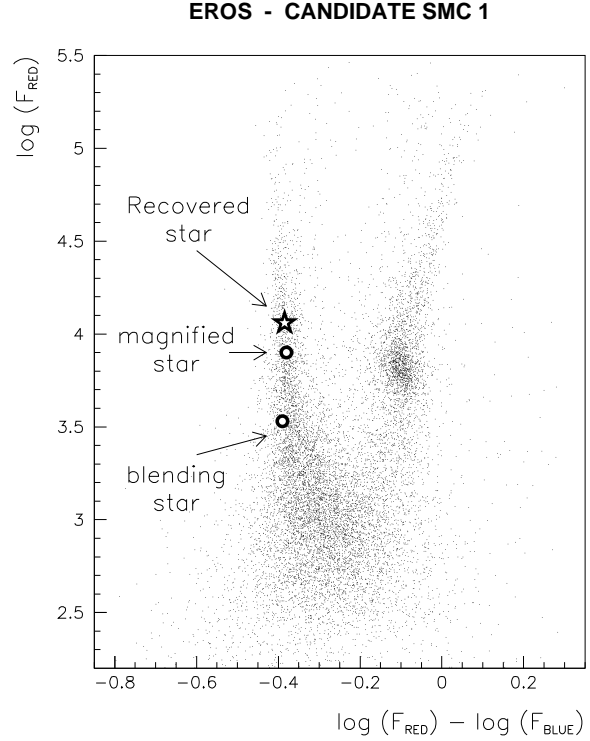,width=8.7cm} \caption{Color-magnitude diagram of
  the field surrounding the microlensing candidate. The exposure time on this
  field was 480~s and is here renormalized to 300~s for comparison with figure
  \ref{cut_hr}. The recovered star is plotted with a star marker, the
  components of the blend with circles.}
  \label{hrcand_1}\end{center}\vspace{-0.2cm}
\end{figure}

Figure \ref{hrcand_1} illustrates the position of the candidate reconstructed
star in the color-magnitude diagram of the surrounding region (star marker), as
well as that of the two components of the blend (circles). They all lie on the
main sequence.

\section{Estimate of optical depth and lens mass}

The optical depth is the instantaneous probability that a given source star be
magnified by more than a factor of 1.34. It can be estimated as 
\begin{equation}
\tau = \frac{1}{N_{\rm obs}T_{\rm obs}}\:\frac{\pi}{2} 
            \sum_{\rm events} \frac{\Delta t}{\epsilon(\Delta t)}
\end{equation}
where $\epsilon(\Delta t)$ is the detection efficiency given in table
\ref{eff}, $T_{\rm obs} = 1$~year and $N_{\rm obs}=5.3\,10^6$ stars. With the
characteristics of the single event described above, this yields (fit with
blending):
\begin{equation}
\tau \simeq 3.3 \times 10^{-7} 
\end{equation}
i.e. about 50\% of the optical depth predicted by a ``standard'' isothermal and
isotropic spherical halo fully comprised of compact objects (cf section 5). It
is consistent with that measured toward the Large Magellanic Cloud
(\cite{Macho2yr}, see also Ansari et al. 1996a).

\begin{figure} [h] %\vspace{-0.1cm} 
  \begin{center} \epsfig{file=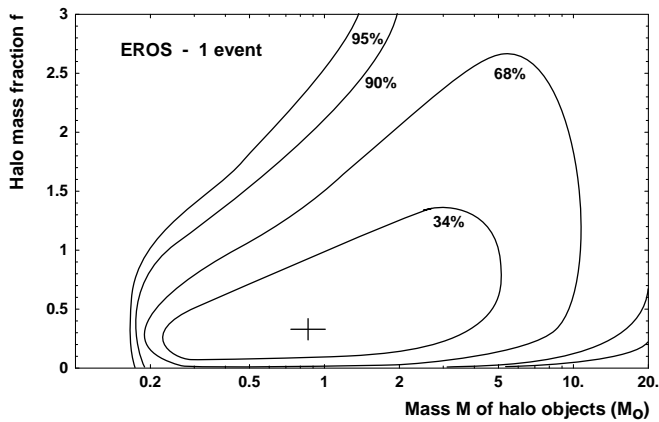,width=8.7cm} \caption{Likelihood
  contours with one microlensing candidate for a standard halo model. The cross
  marks the peak of the 2-D distribution (solution with blending).}
  \label{lkl} \end{center} \vspace{-0.2cm}
\end{figure}

Assuming a standard halo model with a mass fraction $f$ composed of dark
compact objects having a single mass $M$, a likelihood analysis allows us to
estimate the most probable mass of the deflector generating the observed event.
The likelihood is the product of the Poisson probability of detecting $N_{\rm
evt}$ events when expecting $f\,N_M$, by the probability of observing the
time-scales $(\Delta t_1,..\,, \Delta t_{\rm evt})$. We calculate likelihood
contours in the $(\log(M),f)$ plane using a Bayesian method with a uniform
prior probability density in $f$ and in $\log(M)$ (i.e. equal probability per
decade of mass). They are shown in figure \ref{lkl}.
We integrate over $f$ to obtain the 1-D likelihood for the mass of the
deflector. This yields the most probable mass of the Halo deflector, given with
$1\sigma$ error bars:
\begin{equation} 
M = 2.6^{+8.2}_{-2.3} {\rm \;M_\odot} 
\end{equation}
More statistics is obviously required to constrain the mass of halo
deflectors.

\section{Expected number of Halo events --- model dependence}

We studied a wide range of disk-halo models. Table \ref{desc_model} summarizes
their characteristics. The rotation velocity at the Sun predicted by these
models is always within the observational range: $V_{\rm Tot}({\rm
R_\odot})\simeq 200 {\rm \:km/s}$ (Merrifield 1992), or $V_{\rm Tot}({\rm
R_\odot})=220 \pm 15 {\rm \:km/s}$ (Binney and Tremaine 1987). The optical
depth is independent of the mass of the deflectors; the event rate is given
assuming that all deflectors in the halo have a single mass equal to $1{\rm
\;M_\odot}$. To obtain the predicted value of the event rate for other masses,
one only needs to scale by $\eta_M$, the integral of the mass dependence of the
event rate times $f_M(M)$, the normalized mass distribution:
\begin{equation}
 \eta_M = \int f_M(M) \frac{1}{\sqrt{M/{\rm \;M_\odot}}} \:dM 
\end{equation}
This simplifies to $(M/{\rm \;M_\odot})^{-1/2}$ for a Dirac distribution peaked
at $M$.

\begin{figure} [h] \vspace{-.5cm} \begin{center} 
  \epsfig{file=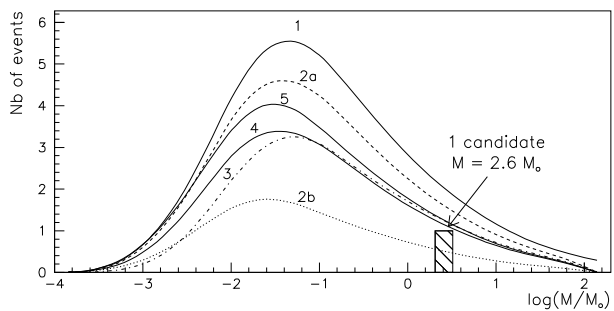,width=8.7cm} \caption{Expected number of events for
  the models described in the text, including the experimental efficiency.}
  \label{nbevt}\end{center}\vspace{-0.2cm}
\end{figure}

\begin{table}[ht]
\begin{center} \begin{tabular}{|c||c|c|c|c|c|c|} 
\hline
{\em MODEL} & 1 & 2a & 2b & 3 & 4 & 5 \\ \hline \hline
$\Sigma_0\:({\rm \;M}_\odot/{\rm pc}^2)$ & 50 & 50 & 100 & 50 & 50 & 80\\
\hline
$\rho_\odot\:({\rm \;M}_\odot/{\rm pc}^3)$ & .008 & .008&.003&.014&.00 &.005 \\ 
$\beta$ & - & 0& 0& 0 & 0.2&0\\ 
$q$ & - & 1 & 1&0.71& 1&1\\ 
\hline \hline
$M({60\:\rm kpc})$ &5.1&1.9&0.7&2.0&1.2&2.2\\
\hline 
$V_{\rm Tot}({\rm R_\odot}) \:({\rm km/s})$ & 192&202&221&205&199&219\\ 
$V_{\rm Halo}(50{\rm\: kpc})$   & 189&164&100&169&134&163\\
$V_{\rm Tot}(50{\rm \:kpc})$    & 199&176&133&180&148&182\\ 
\hline\hline
$\tau\:(10^{-7})$ & 6.8 & 5.7 & 2.1 & 3.9&4.2&3.8\\ 
$\Gamma\:({\rm 10^{-7}\,yr^{-1}})$&22.8&17.8&5.4&14.2&12.6&9.1\\
\hline
\end{tabular}
\caption{Description of the Galaxy models. We give the mass $M$ of the
halos out to the SMC (in units of $10^{11}{\rm \;M}_\odot$), the rotation
velocities (in km/s), the optical depth $\tau$ and the event rate $\Gamma$ for
$1{\rm M_\odot}$ deflectors (with a 100\% efficiency). $\Sigma_0$ is the
central column density, $\rho_\odot$ the local halo density at the Sun,
$\beta$ is proportional to the asymptotic slope of the rotation curve and $q$
is the flattening ratio of the halo.}\label{desc_model}
\end{center} \vspace{-0.5cm}
\end{table}
Model 1 is the ``standard'' halo model: an isotropic and isothermal spherical
halo, with a mass distribution given in spherical coordinates by (Caldwell and
Coulson 1981):
\begin{equation}
\rho(r) = \rho_\odot \frac{R_\odot^2+R_c^2}{r^2+R_c^2} 
\end{equation}
where $R_c=5$~kpc is the Halo ``core radius'' and $R_\odot = 8.5$~kpc is the
distance from the Sun to the Galactic Center. Model 2a is the equivalent
power-law model (Evans, 1993). Model 2b has a maximal disk ($\Sigma_0=100{\rm
\;M}_\odot/{\rm pc}^2$) and a very light halo, model 5 intermediate disk and
halo. Model 3 has a flattened E6 halo (axis ratio $q=0.71$) and model 4 a
decreasing rotation curve ($\beta=0.2$ where $\beta$ is proportional to the
logarithm of the asymptotic slope). Models 2--5 are all power-law halo models,
with self-consistent mass and velocity distributions.  For each of these, we
derive the expected number of events versus the mass $M$ of the objects in the
Halo (assuming a Dirac mass distribution) and compare with the observations (cf
figure \ref{nbevt}).

Whatever the halo model considered, the sole event we observed, if caused by a
deflector in the halo of our Galaxy, corresponds to at least 40\% of the total
optical depth expected for $f=1$. This is a very small number statistics,
however.

\section{Parallax analysis}

The very long time-scale of the observed event suggests that it could show
measurable distortions in its light curve due to the motion of the Earth around
the Sun, (the parallax effect: Gould 1992), provided that the Einstein radius
projected onto the plane of the Earth is not much larger than the Earth orbital
radius. The first detection of parallax in a gravitational microlensing event
was observed by Alcock et al. (1995). The natural parameter to measure the
strength of parallax is the semi major axis of the Earth orbit, $R_O$, in units
of the projected Einstein radius: $\delta u=R_O(1-x)/R_E$ where $x=D_d/D_s$,
with $D_d$ the distance from the observer to the deflector and $D_s$ the
distance to the source.

\begin{figure} [h] 
  \begin{center} \epsfig{file=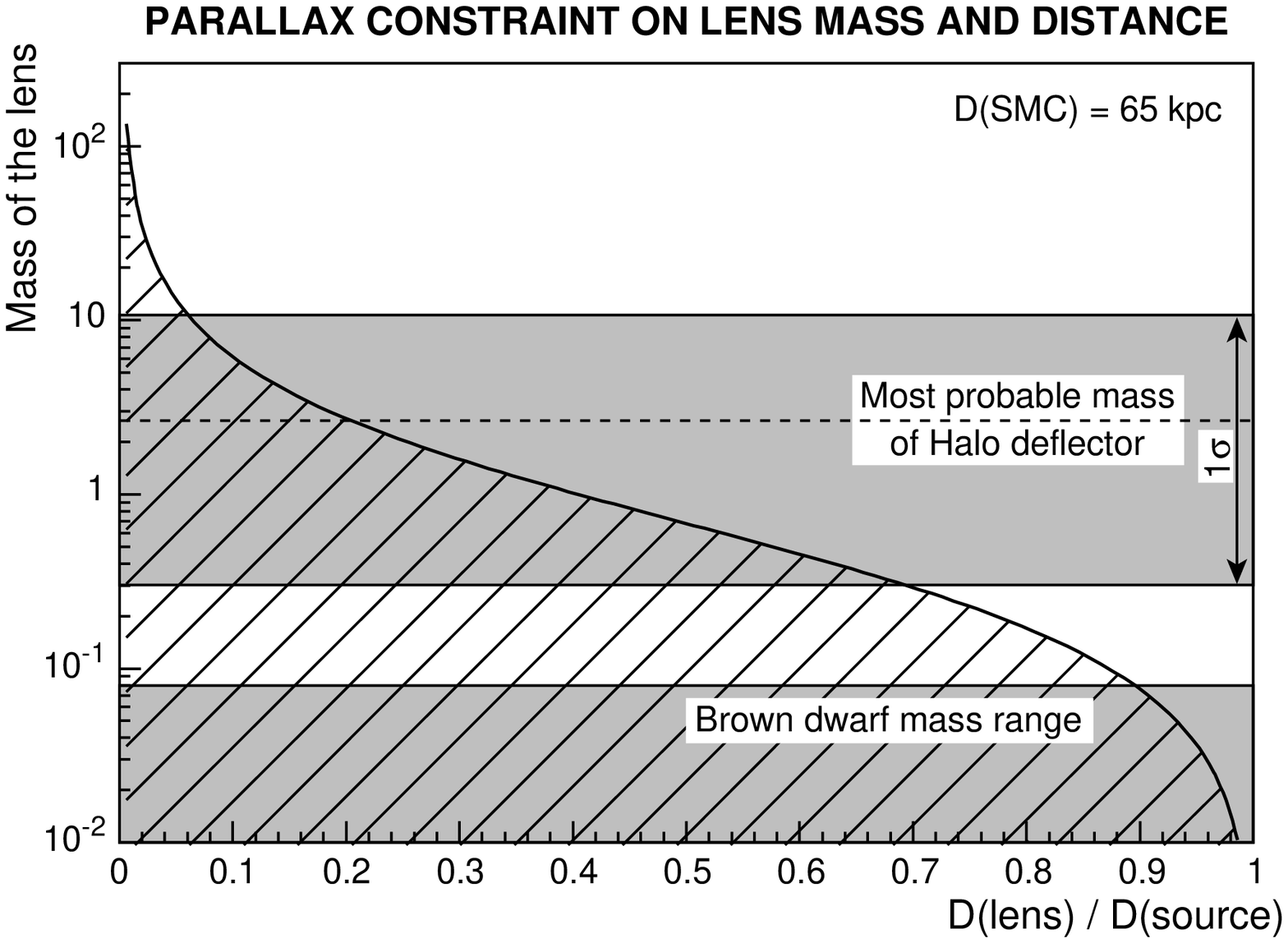,width=8.cm} \caption{Relation
  mass-distance of the deflector, from the parallax analysis. Only the region
  above the curve is allowed. The top gray area is the $1 \sigma$ most probable
  mass of a Halo deflector (see section 4).}  \label{par_mx} \end{center}
%  \vspace{-0.2cm}
\vspace{-0.5cm} \begin{center} \epsfig{file=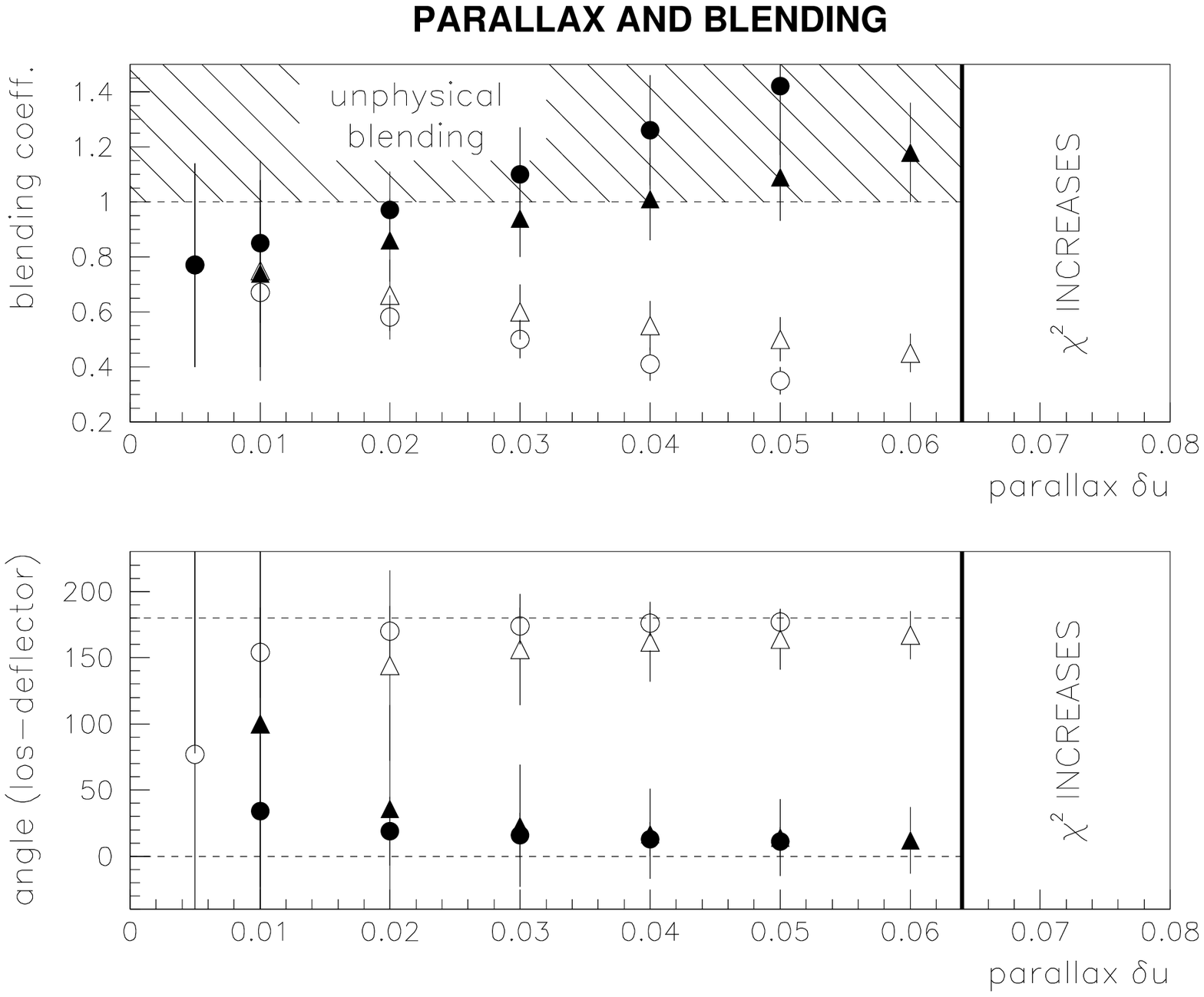,width=8.7cm}
  \caption{Degeneracy valleys between parallax and blending effects. Top plot:
  blending coefficient compensating the effect of parallax. Bottom plot: angle
  (in degrees) between the directions of the line of sight and deflector
  velocities compensating the effect of parallax. Full and empty markers
  correspond to the same $\chi^2/{\rm d.o.f.}$, but the first set tends to an
  angle of 180 degrees, while the second tends to an angle of 0. The four sets
  of markers correspond to the four possible relative motions of the Earth and
  the deflector.}  \label{parbl} \end{center} \vspace{-0.2cm}
\end{figure}

No evidence for distortion due to parallax is detectable on the light curve,
implying either a very massive deflector with a very large Einstein radius, or
a deflector near the source. Because our results for the standard and blended
fits (see table \ref{cand1}) agree well with those obtained by the MACHO
collaboration with a 3 year baseline (Alcock et al. 1997b), we will fix the
level of the baseline flux to that obtained previously with the blended fit, to
perform parallax fits. We fit simultaneously the red and blue light curves
allowing for parallax and for the periodic modulation described in section 3.
Assuming a blending coefficient $c_{\rm bl} = 0.74$, our data allows us to
exclude, at the 95\% CL, that $\delta u > 0.054$. This yields a lower bound on
both the projected transverse velocity of the deflector:
\begin{equation}
\tilde v = R_O / (\Delta t\: \delta u) = v_t/(1-x) > 270 {\rm\; km/s,}
\end{equation}
and on the projected Einstein radius:
\begin{equation}
\tilde R_E = \Delta t\:\tilde v = R_E /(1-x) > 18 {\rm \; AU.}
\end{equation}
We can thus write:
\begin{equation}
(\tilde R_E)^2 = \frac{4GM}{c^2}\times D_s \frac{x}{1-x} > (18{\rm \: AU})^2
\end{equation}
\begin{equation}
\Longrightarrow \frac{M}{\rm M_\odot}\times \frac{x}{1-x} > 0.7
\label{eqpar}\end{equation}
The high projected transverse velocity definitely excludes the possibility that
the deflector is in the disk of the Milky Way, where the typical velocity
dispersion is $\sim 30-40$km/s (Binney and Tremaine 1987). Moreover, a disk
lens (i.e. $x<1/100$) generating this event would have a mass $M>70{\rm
\;M_\odot}$!  For a deflector in the halo, $x<2/3$ at the 95\% confidence level
(for a standard halo) which requires the mass of the deflector $M$ to be at
least $0.3{\rm \;M_\odot}$, while for a deflector in the SMC, if $1-x \simeq
1/10$, the mass of the lens would be $M\simeq 0.07 {\rm \;M_\odot}$. This is
illustrated in figure \ref{par_mx}.

It is possible for some parallax distortions to be largely cancelled out by
blending effects. However, blending distortions of light curves are always
symmetric about the point of highest magnification, while this is not the case,
in general, of parallax distortions. It is only true when the velocity of the
deflector is parallel or anti-parallel to the velocity of the Earth around the
Sun at the moment of highest magnification. Figure \ref{parbl} illustrates the
amount of blending required to compensate the effect of an increasing parallax
while remaining compatible with the observed light curve. All the points
plotted yield a $\chi^2/{\rm d.o.f.}$ for the fit within $1\sigma$ of the
minimum value (157/158). Note the two minima regions in the planes shown in the
figure, one around an angle of 180 degrees between the projected velocities of
the Earth and of the deflector (full markers), while the other (empty markers)
corresponds to a null angle. The shaded area delimits blending coefficients
greater than unity, which is not physical.  As shown in figure \ref{parbl},
$\delta u$ could be larger than 0.054, but only in the unlikely case of
alignment of velocities. We built a likelihood based on the fit with parallax,
blending and a modulation on the magnified star, taking into account the
probability that the alignment of the Earth and deflector velocities were
parallel or anti-parallel. This yields the 95\% CL upper limit $\delta u <
0.06$ (which requires a blending coefficient $c_{\rm bl} = 0.45$). The
projected velocity is then constrained to be at least 190~km/s and equation
\ref{eqpar} becomes $\frac{M}{\rm M_\odot}\times \frac{x}{1-x} > 0.5$.

\section{Discussion --- SMC self lensing}

If the deflector belongs to the Halo of our Galaxy, it is expected to have a
mass greater than a third of a solar mass (see sections 5 and 7); and yet to be
dim enough to avoid direct detection, it could only be a white dwarf, a neutron
star or a black hole. It is also possible, however, that both the lensing
object and the source star belong to the SMC, in which case the deflector would
have a much smaller mass.

\begin{figure} [h] %\vspace{-0.5cm} 
  \begin{center} \epsfig{file=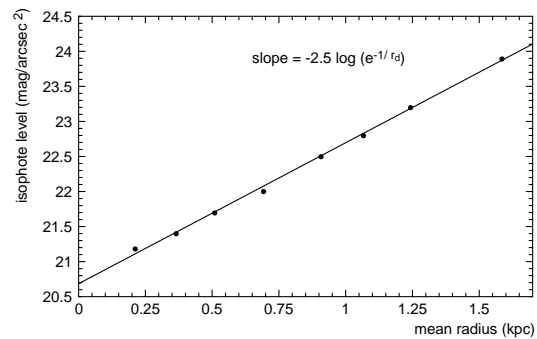,width=7.cm} \caption{Isophote level
  (from the de Vaucouleurs' map) vs. mean distance to optical center of the
  SMC.}  \label{rayonSMC} \end{center} \vspace{-0.2cm}
\end{figure}

Let us estimate the optical depth for SMC self-lensing.  Various authors have
suggested that the SMC is quite elongated along the line-of-sight, with a depth
varying from a few kpc (the tidal radius of the SMC is of the order of 4 kpc)
to as much as 20 kpc, depending on the region under study (Hatzidimitriou and
Hawkins 1989, Caldwell and Coulson 1986, Mathewson et al. 1986). We will
approximate the SMC density profile by a prolate ellipsoid: 
\begin{equation}
\rho = \frac{\Sigma_0}{2h}\:e^{-|z|/h}\:e^{-r/r_{d}}
\end{equation}
where $z$ is along the line-of-sight and $r$ is transverse to the
line-of-sight. The depth $h$ will be a free parameter, allowed to vary between
2.5 and 7.5 kpc. Fitting for instance the Mathewson et al. Cepheid data
(Mathewson et al. 1986) in the bar of the SMC with the above density
distribution gives $h \sim 5.8 \pm 1.2$~kpc (assuming Poissonian statistics on
the Cepheid counts).\footnote{Such a large scale-length in the depth of the SMC
has been criticized, however, by Martin et al. (1990).} The other parameters
are estimated from the surface-brightness map of de Vaucouleurs (see
figure~\ref{carteSMC}) using the identity:
\begin{equation}
R = 26.1 {\rm \:mag/arcsec^{2}} \Longleftrightarrow 1 {{\rm \:L_\odot/pc^2}}
\end{equation} 
derived from $M_V = 4.83 -
2.5\log(L_V/{\rm L}_{\odot,V})$, $M_R-M_V = -0.35$ for the Sun.
Plotting the isophote levels $R$ as a function of the mean distance of each
isophote to the optical center of the SMC (see figure \ref{rayonSMC}), we can
derive the central surface brightness: $R_0 = 20.7{\rm \:mag/arcsec^{2}}$. The
slope of the fit yields the value of the radial scale length: $r_d = 0.54$~kpc.

Assuming a mass-to-light ratio of $\sim 3 {\rm \;M_\odot/L_\odot}$, this gives
a central surface density $\Sigma_0 \simeq 400 {\rm \;M_\odot pc^{-2}}$ and a
total SMC mass of $\sim 1\times 10^9{\rm \;M_\odot}$, compatible with that
estimated from the mass of the LMC considering that the SMC is only $\sim 20\%$
as bright.  We denote as $z_d$ and $z_s$ the positions of the deflector and the
source, both in the SMC, with origin taken at the center of the SMC.

Assuming the same spatial distribution for the source stars and the lenses, the
optical depth for SMC self-lensing is given by
\begin{equation}
\tau = 
\frac{\int \tau(z_s)\;\frac{\rho(z_s)}{M}r\;dr\;d\theta\;dz_s}
     {\int \frac{\rho(z_s)}{M}r\;dr\;d\theta\;dz_s}
\end{equation}
where $\tau(z_s)$ is the usual optical depth due to source stars all at a
distance $z_s$ and deflectors at $z_d$ contained in the elementary volume
$r\:dr\:d\theta\:dz_d$:
\begin{equation}
\tau(z_s)=\int_{-\infty}^{z_s}\frac{\rho(z_d)}{M}\;dz_d\;\pi \frac{4GM}{c^2}
                              \frac{(D_s+z_d)(z_s-z_d)}{(D_s+z_s)}.
\end{equation}
For $h=2.5$, 5.0 or 7.5 kpc, this yields optical depths $\tau = 1.0\;10^{-7}$,
$1.7\;10^{-7}$ or $1.8\;10^{-7}$ respectively. Considering the very limited
statistics we have, this optical depth is consistent with the observations.

Let us also consider the expected typical time-scales of SMC-SMC microlensing
events. The velocity dispersion in the SMC is $<\!\sigma\!>\: \sim$~30 km/s
(Hatzidimitriou et al. 1997, Suntzeff et al. 1986), so the estimated mass $M$
of the deflector causing the observed event ($\Delta t = 123$~days) could be
greatly reduced compared to that of a halo deflector. On average, we have:
\begin{equation}
\frac{M}{\rm M_\odot}\times x(1-x) \simeq 0.0088 
\end{equation}
Thus, if the deflector is 5 kpc (resp. 2.5 kpc) from the source, we have $M\sim
0.1 {\rm \;M_\odot}$ (resp. $0.2 {\rm \;M_\odot}$). 

As more data are accumulated, we expect SMC-SMC events to be highly
concentrated in high density regions of the Cloud (see figure 1), unlike Halo
events which should be distributed like the SMC stars over the sky. In
addition, they should not have measurable parallax distortions. These criteria
will help distinguish between the two possibilities.

\section{Conclusion}

We have presented here the result of a one-year survey toward the Small
Magellanic Cloud with EROS2. One star has a light curve that is best
interpreted as due to microlensing with an Einstein radius crossing time of 123
days when allowing for blending, with 70\% of the total baseline flux
contributed by the star being lensed. The light curve exhibits a 2.5\%
modulation with a period $P\simeq 5.12$~days. The optical depth estimated from
this event is $\sim 3.3 \times 10^{-7}$, to be compared with $\tau = 6.8 \times
10^{-7}$ for a spherical isothermal halo containing only such objects. The most
probable mass of the deflector (if it is in the Halo) would be $M =
2.6^{+8.2}_{-2.3} {\rm \;M_{\odot}}$. If we interpret this event as due to a
deflector in the SMC, the expected optical depth is $\tau \simeq 1.7 \times
10^{-7}$ for a depth scale-length of the SMC of 5~kpc, and the mass of the
deflector would be reduced to about $0.1 {\rm \;M_{\odot}}$. Given the very
small statistics, the SMC interpretation seems possible. Furthermore, we
observe no distortion on the light curve due to the varying velocity of the
Earth on its orbit around the Sun (parallax), although some would have been
expected for such a long duration event, unless the deflector were either very
heavy or near the source. This further supports the SMC lens interpretation.

Further observations will help discriminate Halo from Cloud deflectors. In
particular, because the velocity dispersions in the LMC and the SMC differ by
almost a factor of 2, the observation of a significant trend for longer
time-scale events toward the SMC than toward the LMC would be a clear signature
of events dominated by SMC or LMC self-lensing.

\begin{acknowledgements}
We are grateful to D. Lacroix and the technical staff at the Observatoire de
Haute Provence and to A. Baranne for their help in refurbishing the MARLY
telescope and remounting it in La Silla. We are also grateful for the support
given to our project by the technical staff at ESO, La Silla. We thank
J.F. Lecointe for assistance with the online computing. We also thank the staff
of the CCIN2P3, and in particular J. Furet, for constant help with the mass
production of the data. We thank the referee for the many improvements he 
suggested.
\end{acknowledgements}

\begin{noteadd}
After submission of this article, the OGLE-2 collaboration confirmed, with
their data taken after June 1997, the modulation we detected. They also
separated the two components of the blend, which allowed them to establish that
the modulation affects the magnified component (Udalski et
al. astro-ph/9710365). The MACHO collaboration provided us their data on the
candidate, which allowed us to confirm the period, $P=5.124\pm 0.001$ days, and
the amplitude, $(2.4\pm 0.4)$\%, of the modulation; we thank them for this
private communication.

\end{noteadd}

\end{document}